\documentclass[useAMS,usenatbib]{mn2e}

\usepackage{graphicx}
\newcommand{\msun}{M$_{\odot}$}
\newcommand{\kms}{$km\,s^{-1}$}

\title{Mass Loss in 2D ZAMS Stellar Models}

\author[C.C. Lovekin]{C.C.Lovekin$^1$\thanks{email: clovekin@lanl.gov}\\
$^1$T-2, Los Alamos National Laboratory, Los Alamos, NM, 87545}

\begin{document}

\date{Accepted , Received }
\pagerange{\pageref{firstpage}--\pageref{lastpage}}\pubyear{2011}

\maketitle

\label{firstpage}

\begin{abstract}
A large number of massive stars are known to rotate rapidly, resulting in significant distortion and variation in surface temperature from the pole to the equator.  Radiatively driven mass loss is temperature dependent, so rapid rotation produces variation in mass loss and angular momentum loss rates across the surface of the star, which is expected to affect the evolution of rapidly rotating massive stars.  
In this work, we use ZAMS stellar models to investigate the two dimensional effects of rotation on stellar mass loss, using two common prescriptions for radiatively driven mass loss.  The associated loss of angular momentum from these models is also considered.    
Using 2D stellar models, which give the variation in surface parameters as a function of co-latitude, we implement two different mass loss prescriptions describing radiatively driven mass loss. 
We find a significant variation in mass loss rates and angular momentum loss as a function of co-latitude.  We find that the mass loss rate decreases as the rotation rate increases for models at constant initial mass, and derive scaling relations based on these models.  When comparing 2D to 1D mass loss rates, we find that although the total angle integrated mass loss does not differ significantly, the 2D models predict less mass loss from the equator and more mass loss fr om the pole than the 1D predictions using von Zeipel's law.  As a result, rotating models lose less angular momentum in 2D than in 1D, which will change the subsequent evolution of the star.  The evolution of these models will be investigated in future work.

\end{abstract}


\section{Introduction}
\label{intro}

Radiatively driven mass loss plays an important role in determining the properties of the circumstellar environment, and is also expected to have a significant impact on the subsequent evolution of the star itself.  Predominantly important in massive stars, the radiatively driven stellar winds are an important factor in determining the course of evolution and final end state of the star.  However, many massive stars are known to be rapidly rotating, which produces variation in the flux and effective temperature over the stellar surface.  This variation in turn, will affect the temperature dependent mass loss rates, producing latitudinally dependent stellar winds \cite[see, for example][]{owocki98,dwarkadas2002}.  The effects of rotation on stellar mass loss rates can greatly influence the later evolution of a star through angular momentum loss \citep{meynet00}.  It is therefore important for us to understand how exactly mass loss rates depend on rotation, and the subsequent effects on evolution.    

In general, the effects of stellar rotation on wind profiles are not well understood.  Research into this problem is still in its infancy, and the results can be highly contested.  Several studies have shown that at very high rotation rates, the wind can switch to a slower, denser equatorial outflow \citep{pelupessy2000,cure04,cr04,cure05,madura07}, which may explain the observed disks around B[e] stars.  However, other groups find that when the effects are modelled in 2D, the opposite is true, and the wind density is higher at the pole than the equator \citep{petrenz2000}.  However, in this work, we focus instead on how rotation affects the mass loss rates from the surface of the star, and the implications these mass loss rates have for the structure and evolution of the star, rather than the winds.  

Early investigations into latitudinally dependent mass loss suggested that the reduction in effective gravity near the equator would produce an enhancement in the equatorial density \citep{friend86}.  The increase in mass lost at the equator would carry away more angular momentum, and the star would spin down faster.  However, these results assumed the flux to be constant across the surface of the star.  When the effects of gravity darkening are taken into account, (using, e.g., von Zeipel's law,) the opposite was found to be true, and the star was found to lose most of it's mass from the polar regions \citep{dwarkadas2002}.  In this case, the star will spin down more slowly, which could lead to relatively rapidly rotating stars in later phases of evolution, such as the luminous blue variable (LBV) or Wolf-Rayet (WR) phases.  

A change in the mass loss rate will change the amount of angular momentum lost from a star, causing it to spin down differently.  As discussed by \citet{meynet00}, massive stars can lose mass and angular momentum quickly enough to prevent the star reaching its critical rotation velocity on the main sequence.  As a result, lower mass stars with lower mass loss rates would end their main sequence lives as more rapid rotators than higher mass stars.  At the end of the main sequence, low mass rapidly rotating stars may reach the critical velocity as angular momentum is dredged up from the core.  At the critical velocity, these stars would lose significant amounts of mass from the equator.  This could explain why extremely rapidly rotating stars, such as Be stars are not seen among the more massive O stars \citep{meynet00}.  The interaction between mass loss and angular momentum would also explain the relative frequency of Be stars in the LMC and SMC.  At lower metallicity, the mass loss rates are typically lower, so stars in the LMC and SMC may be able to continue to rotate faster, and are hence more likely to form Be stars \citep{meynet00}.  

A recent study by \citet{georgy} considers the effects of anisotropic winds on the evolution of a rotating star.  Using the Roche approximation to determine the shape of the surface equipotential, they find that as the rotation rate increases, more mass is lost from the pole.  At high rotation rates, the difference in flux between the pole and equator becomes quite large, with the flux at the pole more than twice the flux at the equator when the rotation rate is greater than 0.8 $\Omega_c$.  Using a semi-analytical approach, when anisotropic effects are taken into account, the angular momentum loss changes by less than 4 \% compared to isotropic winds for rotation rates up to 0.9 $\Omega_c$, and the subsequent effects on the stellar evolution are expected to be small. Using more realistic mass loss rates there is basically no change in the angular momentum loss until $X_c$ $\sim$ 0.4.  At this point, the star becomes sufficiently rapidly rotating to significantly reduce the temperature of the equatorial regions, which produces a strong equatorial mass loss in the anisotropic case \citep{georgy}. 

An understanding of the latitudinal distribution of mass loss is also important for understanding the structure of the surrounding nebulae.  The shape of the circumstellar nebulae may also have an effect on supernovae light curves \cite{vanmarle}.  Early studies suggested that if the mass loss in a rotating star is primarily radial, the material carried away could lead to the formation of a wind compressed disk \citep[WCD,][]{bjorkman}.  However, more recent research including the non-radial components of the wind have shown that this flow inhibits the formation of a WCD, and may be more likely to produce a polar nebulae \citep{owocki98}.  The effects of rotation on stellar winds are still very uncertain, with simulations finding polar density enhancements \citep{petrenz2000} or equatorial density enhancements \citep{pelupessy2000,madura07}.  Different mechanisms may be important in different stages of stellar evolution, as observations of B[e] stars are thought to be explained by a dense equatorial wind \citep{zick85,zick89}, while nebulae around LBVs are often bipolar, as is seen around $\eta$ Carinae.  Asymmetric mass loss produced by rotation may be able to explain the structure of some of these nebulae.

Global effects of rotation on mass loss have been investigated by several authors.  \citet{friend86} found that the mass loss rate of a star at a given location in the HR diagram increases with increasing rotation rate.  This effect has been  confirmed observationally \citep{vardya,nieuwenhuijzen}, and has also been found in models by \citet{maeder99}.  This relationship between mass loss and rotation rate can be described by a simple equation \citep[eg,][]{bjorkman,langer}:
\begin{equation}
\label{eqn:1Dscale}
\frac{\dot{M}(v)}{\dot{M}(v_{rot} = 0)} = \left(\frac{1}{1-v}\right)^{\xi}
\end{equation}
for a star at constant temperature and luminosity.  The exponent $\xi$ is usually taken to be $\sim$ 0.43.  According to this equation, the enhancement of the mass loss rate due to rotation is modest, only a factor of 2 for $\Omega$ = 0.8.  Formally, this formula leads to infinite mass loss rates at critical rotation velocities, so this scaling law breaks down at very high rotation rates.  It is this formula that is used to calculate the effects of rotation on mass loss in many stellar evolution codes \citep[eg,][]{langer, maeder00}. 

The rotational enhancement of mass loss described above can be even more important in stars that are already close to the Eddington limit.  In this case, for a star rotating at a given angular velocity $\Omega$, Equation \ref{eqn:1Dscale} can be rewritten to include the effects of the Eddington luminosity as:
\begin{equation}
\label{eqn:eddington}
\frac{\dot{M}(\Omega)}{\dot{M}(v_{rot} = 0)} = \frac{(1-\Gamma)^{\frac{1}{\alpha}-1}}{\left[1-\frac{\Omega^2}{2\pi G \rho_m}-\Gamma\right]^{\frac{1}{\alpha}-1}}
\end{equation}
where $\rho_m$ is the average density inside a given surface, and $\alpha$ is one of the force multiplier parameters.  In this case, the mass loss rate depends on both the rotation rate and the Eddington factor, a situation known as the $\Omega\Gamma$ limit \citep{maeder00}.  In extreme cases, with $\Gamma > 0.639$, even moderate rotation can cause the denominator to vanish, leading to infinite mass loss rates. \citet{maeder00} show that the ratio $\dot{M}(\Omega)/\dot{M}(0)$ can be quite moderate for stars up to 40 \msun, but can become large for stars above 60 \msun, and can start to diverge for stars close to the Humphreys-Davidson limit.    

Despite recent progress, there is still much to be discovered on the interaction between stellar winds and rotation.  Scaling laws, such as those used in Equations \ref{eqn:1Dscale} and \ref{eqn:eddington} are derived for stars at constant temperature and luminosity, but both of these quantities change as the rotation rate increases.  Current 1D and pseudo-2D models do not fully account for the distortion of the surface and variation in temperature, making realistic calculations of the anisotropy in the mass loss and angular momentum loss rates suspect.  A more realistic calculation of these quantities could produce significant discrepancies compared to current stellar evolution models.   

In this work, we calculate the stellar mass loss rates for fully 2D stellar structure calculations, addressing many of the issues discussed above.  We consider two different mass loss prescriptions, and compare the 2D results to 1D calculations and scaling laws.  In this way, we can directly calculate the distribution of mass loss, and its effect on angular momentum loss and the evolution of the star.  

In Section \ref{sec:models}, we discuss the 2D stellar structure models used in this work.  In Section \ref{results} we compare our 2D calculations to the mass loss rates and angular momentum loss rates in 1D models.  In Section \ref{rates} we discuss the effect of rotation on the global mass loss rates of stellar models, and present scaling relations derived at constant mass.  Finally, we summarize our results and discuss future work in Section \ref{conclusions}.  

\section{Models}
\label{sec:models}
\subsection{Stellar structure}

The stellar models used here were calculated using the 2D stellar structure code {\tt ROTORC} \citep{bob90,bob95}.  The code uses the OPAL opacities \citep{opal96} and equation of state \citep{opaleos}.  In this work, we consider ZAMS models of 20, 30 and 40 \msun, with X = 0.7 and Z = 0.02.  These models solve the conservation equations of mass, momentum, energy and hydrogen abundance, along with Poisson's equation for the gravitational potential on a two-dimensional finite difference grid with the fractional surface equatorial radius and the colatitude as the independent variables.  This grid contains 581 radial zones and 10 angular zones.  Simultaneously with the conservation equations, we solve the integral of the density over the model, and set the surface equatorial radius such that the result equals the total mass of the model.  The surface is assumed to be an equipotential determined by the equatorial gravitational potential.  The models, summarized in Table \ref{models}, are uniformly rotating ZAMS models, with velocities equal to 0, 0.3, 0.5 and 0.7 $\Omega_c$.  

\begin{table*}
\centering

\caption{\label{models}Model properties}
\begin{tabular}{cccccccc}
\hline
Mass & V$_{eq}$ & $\Omega/\Omega_c$ & R$_{eq}$ & R$_p$/R$_{eq}$ & T$_{eff}$ & $\Delta T$ \footnote{Temperature difference between pole and equator}& L/L$_{\odot}$\\
(\msun) & (km $^{-1}$) &  & (R$_{\odot}$) &  & (K) & (K) & \\

\hline

20 & 0 & 0 & 5.835 & 1.000 & 34476 & 0 & 42899 \\
20 & 200 & 0.3 & 5.991 & 0.969 & 34090 & 1161 & 42313 \\
20 & 375 & 0.5 & 6.437 & 0.892 & 33168 & 3866 & 41196 \\
20 & 550 & 0.7 & 7.376 & 0.770 & 31802 & 7899 & 40122 \\
30 & 0 &  0 & 7.310 & 1.000 & 39598 & 0 & 117162 \\
30 & 225 & 0.3 & 7.528 & 0.967 & 39195 & 1470 & 116564 \\
30& 400 & 0.5 & 8.038 & 0.897 & 38203 & 4263 & 113517 \\
40 & 0 & 0 & 8.621 & 1.000 & 42959 & 0 & 225732 \\
40 & 250 & 0.3 & 8.893 & 0.964 & 42395 & 1791 & 222164\\
40 & 425 & 0.5 & 9.487 & 0.897 & 41336 & 4711 & 216740\

\end{tabular}
\end{table*}

\subsection{Including mass loss}
\label{massloss}

As {\tt ROTORC} gives us the values of radius and effective temperature as a function of co-latitude, we can use these to calculate the local mass loss rate as a function of co-latitude in a rotating star.  This has been done using two mass loss prescriptions:  the numerical mass loss rates of \citet{vink01}, hereafter V01, the original theoretical formulation for radiatively driven winds of \citet{cak} (CAK) including the finite disk correction, described by \citet{kud}, hereafter K89.  These two rates allow us to compare a theoretical formulation (K89) and the numerically derived rates commonly used in stellar evolution codes (V01).

In each angular zone, the local radius and effective temperature are calculated.  These are used to calculate a local luminosity, which is defined as the luminosity of a spherical star with the given radius and the effective temperature.  Using these parameters, we can calculate a local Eddington factor, defined as 
\begin{equation}
\Gamma_e = \frac{L(\theta)\sigma_e(\theta)}{4\pi c G M}
\end{equation}
where $\sigma_e$ is the electron scattering cross-section, which depends on the ionization fraction (and hence temperature) and is hence allowed to vary as a function of co-latitude.  

If the V01 mass loss rates are selected, the local Eddington factor is used to calculate the temperature of the bi-stability jump.  The mass loss rate is then calculated in each angular zone using Equation 24 of V01 if the local effective temperature is above the temperature of the bi-stability jump, and their Equation 25 if the temperature is below the bi-stability jump.  Below 12500 K, the V01 mass loss rates are undefined, and we neglect the mass loss from these regions.  For the models considered here, all of the surface temperatures are greater than 25000 K, so this cutoff has no effect in our models.

If the K89 mass loss rates are selected,  the force multiplier parameters ($\alpha$, $\delta$ and $k$) must also be determined locally.  All of these parameters are temperature dependent, and we have used a parametrization based on the tabular values given in \citet{leitherer10}.   For $k$, we used data taken from their Tables 1 and 2 for solar metallicity stars of luminosity class V, and fit using a least squares analysis.  This fit was implemented in our mass loss code, allowing $k$ to be determined locally based on the effective temperature and luminosity.  The values of $\alpha$ and $\delta$ were set as a step function, with ($\alpha,\delta$) = (0.598,0.05) for $T_{eff} > 30000K$ and (0.607, 0.07) below this.  For the K89 rates, we followed the method for calculating the mass loss rates outlined in K89, section 6, but we allow the luminosity, effective temperature, and the force multipliers $\alpha$, $\delta$ and $k$ to be functions of co-latitude.  

The force multiplier parameter description of stellar winds is likely an over-simplification, and has been challenged by (for example) \citet{lucy07, meuller08} based on Monte Carlo simulations.    A more detailed parametrization for the force multiplier has been described by \citet{kud02}.  However, this new parameterization is much more complicated and must be solved iteratively.  In the interest of computational efficiency, we have chosen to use the original force multiplier parameter description.  

Mass loss rates are typically given in terms of \msun$yr^{-1}$, but this also has an implicit unit of per star.  Since we are calculating \emph{local} mass loss rates, we must scale the calculated mass loss rates by the area of the zone.   The calculated mass loss rates are multiplied by the time step to determine the total amount of mass lost, and this is then multiplied by the ratio of the area of the zone to the total area of a spherical star with the local radius to give the amount of mass lost from this zone.  At each co-latitude, the mass is lost by removing surface zones until the required amount of mass has been lost.  Finally, the total mass of the star is reduced by the amount of mass lost at this time step.  The star is then allowed to relax back to an equilibrium configuration.  

As well as calculating mass loss in the zone-by-zone fashion described above, the code can also calculate a pseudo-1D mass loss rate, based on the luminosity and global effective temperature of the star.  Calculating a pseudo-1D rate in this way allows us to make a more direct comparison with other 1D calculations.  We use the luminosity and  global effective temperature to calculate a single mass loss rate for the model.  The total mass lost in this way is then divided among the angular zones weighted by area, the mass of the star is modified and the star is allowed to relax as described above.  

The evolution of a rotating star including mass loss proceeds as follows.  First, the star is allowed to evolve for some period of time.  To lose mass, the evolution is halted and mass is lost as described above.  Once the star has returned to an equilibrium configuration, the evolution is restarted, and continues until the next mass loss step.  This process is repeated until the star reaches the desired evolutionary stage.  In this paper, however, we will consider only zero-age main sequence (ZAMS) models.  The effects of 2D mass loss on stellar evolution will be discussed in a future paper.  

\section{Results}
\label{results}

\subsection{Mass Loss Rates}
\label{massresults}

\begin{figure}
\includegraphics[width=\linewidth]{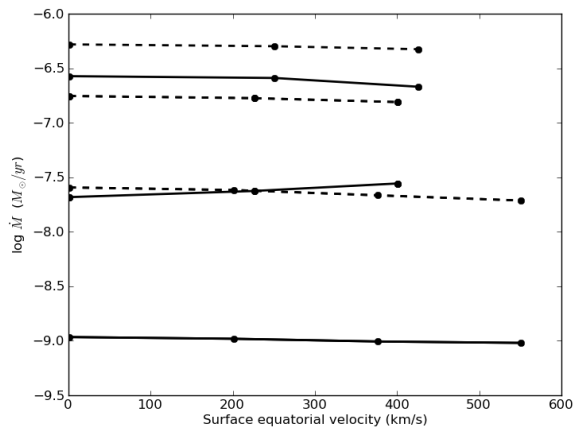}
\caption{\label{fig:vrates}The mass loss rates as a function of rotation rate for the K89 rates (solid) and the V01 rates (dashed).  From bottom to top, the rates are shown for 20, 30 and 40 \msun\ models.  With the exception of the Kudritzki rates at 30 \msun, the mass loss rates all decrease as the rotation rate increases.}
\end{figure}

The effects of rotation on mass loss rates can be seen in Figure \ref{fig:vrates} for the two considered mass loss rates.  The V01 rates are always higher than the K89 rates, but the difference decreases as the mass of the model increases.  This is consistent with the results of \citet{vink00}, who found that their theoretical rates resolved the discrepancy between theoretical and observed mass loss rates for normal O stars.  The observed rates are typically a factor of two higher than those predicted by radiation-driven wind theory, but the \citet{vink00} and V01 rates are in good agreement with observations for these stars.  

In general, the mass loss  rates decrease as the rotation rate increases.  The exception is the K89 rate in the 30 \msun\ models, which increase slightly with increasing rotation rate.  In this model, the effective temperature is close to 40\,000 K, where our parametrization of $k$ has a discontinuity due to lines from CNO elements.  Above 40\,000 K, $k$ increases sharply, and the resulting mass loss rates are higher than at temperatures below 40\,000 K.  In the 30 \msun\ models, the effective temperatures are around 38\,000-39\,000 K, just below this gap.  However, as the rotation rate increases, the pole becomes hotter, and temperatures in this region are above 40\,000 K.  This local increase in the mass loss rate at the pole increases the overall mass loss rate for the star.  The effect becomes more pronounced as the rotation rate increases, and the mass loss rate increases with rotation.  The 20 and 40 \msun\ models are sufficiently far from this jump that the mass loss rates are not affected.  The V01 models do not depend on our parameterization of the force multiplier parameters, and so this effect is not seen in these rates.  

\begin{figure}
\includegraphics[width=\linewidth]{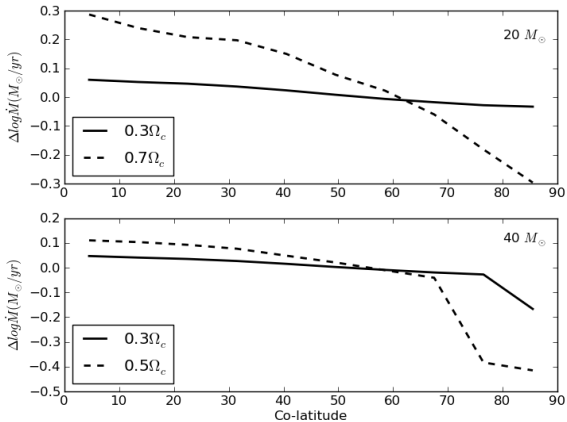}
\caption{\label{fig:kudrates}The variation with colatitude in the mass loss rates for the 20 \msun\ model (top) and the 40 \msun\ model (bottom).  The log of the ratio of the 2D rates to the pseudo-1D rates are shown for the K89 mass loss rates at 0.3 $\Omega_c$ (solid) and at 0.7 $\Omega_c$ and 0.5 $\Omega_c$ (dashed) for the 20 and 40 \msun\ models respectively.  Although the total mass lost is similar in the pseudo-1D and 2D calculations, more mass is lost from the pole in the 2D case, even at relatively slow rotation rates.}
\end{figure}

\begin{figure}
\includegraphics[width=\linewidth]{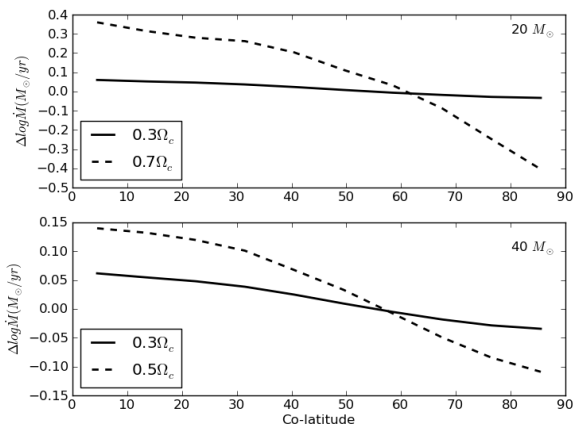}
\caption{\label{fig:vinkrates}As for Figure \ref{fig:kudrates}, but using the V01 mass loss rates.}
\end{figure}

Using our 2D models, we can also compare the anisotropy of the mass loss rates to that predicted using our pseudo-1D models.  The difference between 2D and pseudo-1D calculations is shown for the K89 rates in Figure \ref{fig:kudrates} and for the V01 rates in Figure \ref{fig:vinkrates}.  As expected, our 2D models lose more mass at the pole and less mass from the equator than the pseudo-1D models.  The effects can be significant (nearly 0.1 dex at the pole) even for relatively slowly rotating models (0.3 $\Omega_c$).  In Figure \ref{fig:kudrates}, the effect of the jump in $k$ near 40\,000 can be seen in the sharp decrease in mass loss rate near the equator of the 40 \msun\ model.  Compare this with the 20 \msun\ model, where the temperature is always well below the jump, and the mass loss rate varies smoothly with co-latitude.  

Despite the differences in distribution of mass loss, the total integrated mass lost in the 2D and pseudo-1D calculations is nearly identical.  The total mass lost in each case is shown in Table \ref{tab:massloss}.  The difference in integrated mass lost is largest for models where the range of surface temperatures covers the jump in mass loss rates caused by $k$ (30 and 40 \msun\ models with the K89 rates). The larger the temperature difference between the pole and the equator, the more zones are affected by the jump, and the bigger the difference in mass loss between the pseudo-1D and 2D models.  With the V01 rates, which are not affected by the jump in $k$, the difference in total mass lost is typically less than 2 \%, except in the most rapidly rotating models.   In later phases of evolution, when the temperature of the star drops to the temperature of the bi-stability jump (22\,500-26\,000K), a similar result can be expected in the V01 mass loss rates.

\begin{table*}
\caption{\label{tab:massloss}Total mass (\msun) lost over 1000 years}
\begin{tabular}{cccccccc}
\hline
 &  &  & K89 & &  & V01 & \\
Mass & $v_{eq}$ & pseudo-1D & 2D & \% difference & pseudo-1D & 2D & \% difference\\
20 & 200 & 1.0463e-6 & 1.0436e-6 & 0.26 & 2.4137e-5 & 2.4148e-5 & 0.045 \\
20 & 375 & 9.7383e-7 & 9.8561e-7 & 1.2    & 2.1206e-5 & 2.1598e-5 & 1.8  \\
20 & 550 & 9.1649e-7 & 9.5660e-7 & 4.2    & 1.7708e-5 & 1.9337e-5 & 8.4 \\
30 & 225 & 1.9915e-5 & 2.3663e-5 & 16     & 1.6802e-4 & 1.6802e-4 & 0.0012 \\
30 & 400 & 1.8372e-5 & 2.7743e-5 & 34     & 1.5354e-4 & 1.5473e-4 & 0.77 \\
40 & 250 & 2.6088e-4 & 2.5734e-4 & 1.4    & 5.0327e-4 & 5.0317e-4 & 0.020 \\
40 & 425 & 2.4987e-4 & 2.1379e-4 & 17     & 4.7070e-4 & 4.7230e-4 & 0.34 \\
\end{tabular}
\end{table*}

\subsection{Angular Momentum Loss}
\label{angmom}

\begin{figure}
\includegraphics[width=\linewidth]{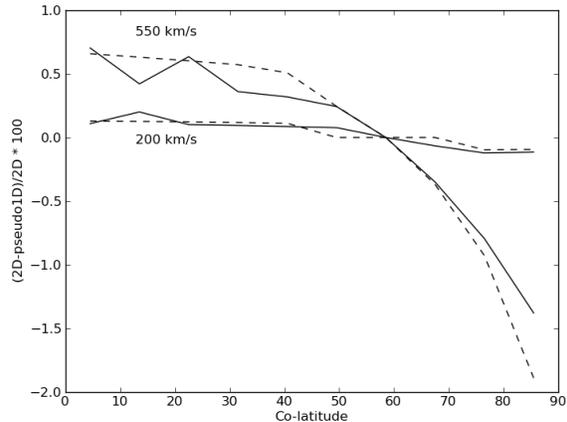}
\caption{\label{fig:angloss}Percent difference in angular momentum loss rate for the 2D and pseudo-1D mass loss rates in the 20 \msun\ model.  Solid lines show the results using the K89 mass loss rates, while dashed lines show the results for the V01 rates.  Shown are curves for models rotating at 550 \kms\ and 200 \kms.}
\end{figure}

Our technique for modelling the mass loss in 2D also allows us to directly determine not only the total rate of angular momentum loss, but also the rate of angular momentum loss from each surface zone.  This can be done for both the pseudo-1D and the 2D mass loss rates described in section \ref{massloss}.  The surface variation in angular momentum loss is shown in Figure \ref{fig:angloss} for the 20 \msun\ models rotating at 200 and 550 \kms.  The overall rate of angular momentum loss predicted by the pseudo-1D mass loss is higher than that predicted by the fully 2D mass loss rates, as shown in Table \ref{tab:angmomloss}.  The 2D model therefore loses less angular momentum through radiatively driven winds, and hence spin down more slowly than the 1D models.  As angular momentum is transported from the core to the surface, a slower rate of angular momentum loss will result in either more rapidly rotating stars at the end of the main sequence, or more mass lost at the equator as the star reaches its critical velocity.  

\begin{table*}
\caption{\label{tab:angmomloss}Total angular momentum lost over 1000 years}
\begin{tabular}{cccccccc}

\hline
 &  &  & K89 &  &  & V01 & \\
Mass & $v_{eq}$ & pseudo-1D & 2D & \% difference & pseudo-1D & 2D & \% difference\\
\hline
20 & 200 & 3.06096e47 & 2.93396e47 & 4.3 & 6.35610e48 & 6.19461e48 & 2.6 \\
20 & 375 & 5.54387e47 & 4.90389e47 & 13  & 1.10423e49 & 9.51412e48 & 16  \\
20 & 550 & 8.17646e47 & 5.93988e47 &  38 & 1.39251e49 & 1.00108e49 &  39 \\
30 & 225 & 7.57032e48 & 7.11083e48 &  6.5& 6.11304e49 & 5.77419e49 & 5.9 \\
30 & 400 & 1.25980e49 & 1.19815e49 & 5.1 & 1.02423e50 & 8.76453e49 & 17 \\
40 & 250 & 1.16000e50 & 1.11368e50 & 4.2 & 2.25489e50 & 2.15184e50 & 4.8 \\
40 & 425 & 1.89198e50 & 1.33934e50 & 41  & 3.48765e50 & 3.06850e50 & 14 \\
\hline
\end{tabular}
\end{table*}

\begin{figure}
\includegraphics[width=\linewidth]{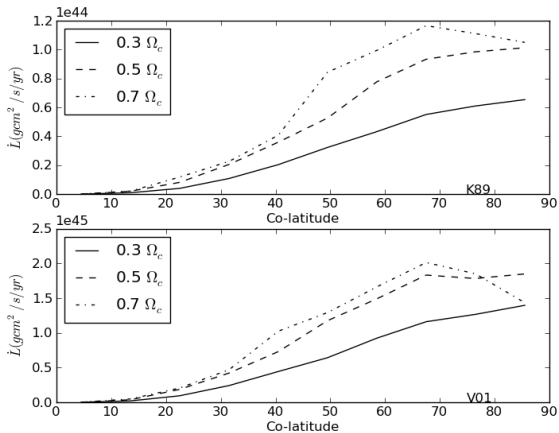}
\caption{\label{fig:angrate}Rate of angular momentum loss calculated in 2D for a 20 \msun\ model using the K89 (top) and V01 (bottom) mass loss rates.  The three curves show the variation in the rate of angular momentum loss across the surface for models rotating at 0.3 $\Omega_c$ (solid), 0.5 $\Omega_c$ (dashed), and 0.7 $\Omega_c$ (dot-dashed).}
\end{figure}

\begin{figure}
\includegraphics[width=\linewidth]{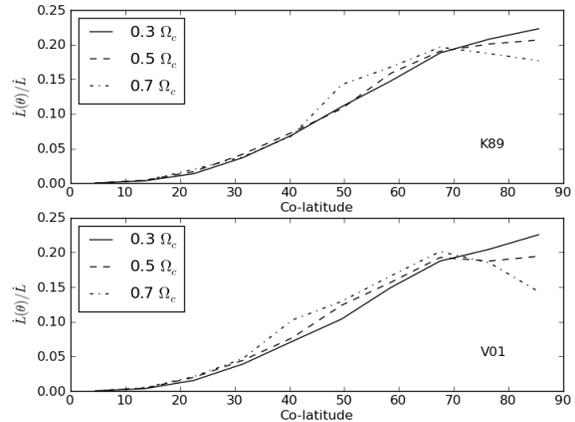}
\caption{\label{fig:scaled}Local rate of angular momentum loss years scaled by the total angular momentum loss of the model for 20 \msun\ models rotating at 0.3 $\Omega_c$ (solid), 0.5 $\Omega_c$ (dashed), and 0.7 $\Omega_c$ (dot-dashed).  Angular momentum loss was calculated using the K89 (top) and V01 (bottom) mass loss rates.}
\end{figure}

As expected, as the rotation rate of a model increases, the rate of angular momentum loss increases as well, as illustrated in Figure \ref{fig:angrate}.  However, at higher rotation rates the total angular momentum of the models also increases.  For each model and mass loss rate, we calculated the ratio of the local rate of angular momentum loss to the total angular momentum loss ($\dot{L}(\theta)/\dot{L}$).  For a given model, this quantity is relatively constant as rotation rate increases, as shown in Figure \ref{fig:scaled} for the 20 \msun\ model.  This is true for any given model, but different masses and mass loss prescriptions scale differently, making it difficult to derive a simple scaling law to describe the angular momentum loss.  In more rapidly rotating and more massive models, there is more variation in mass loss near the equator, so the fractional rate of angular momentum loss does decrease as the rotation rate increases.  This is particularly true for the V01 models, while the K89 models show a more consistent scaling.

\section{Discussion}
\label{rates}

\subsection{Global mass loss}

The current paradigm for stellar mass loss uses Equation \ref{eqn:1Dscale} to scale mass loss rates with stellar rotation for stars at constant luminosity and effective temperature.   As discussed in Section \ref{intro}, this effect has been  seen observationally \citep{vardya,nieuwenhuijzen}, although the magnitude of the effect appears to be somewhat uncertain, ranging from a few tens of percent to two orders of magnitude.  However, stellar models use mass as the independent variable, and the luminosity and effective temperature are allowed to vary.

\begin{figure}
\includegraphics[width=\linewidth]{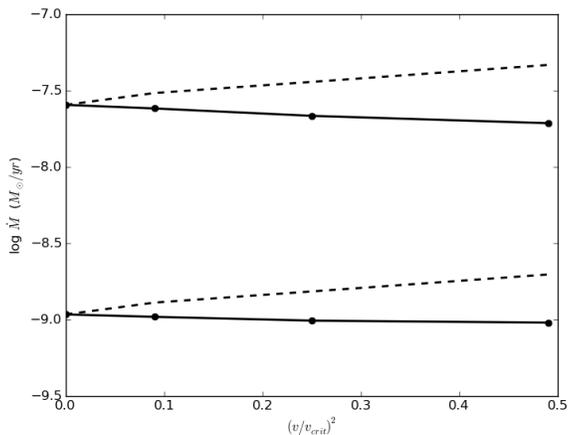}
\caption{\label{time}The 2D mass loss rates for a 20 \msun\ model as a function of rotation rate (solid curves) for the K89 (bottom) and V01 (top) mass loss prescriptions.  The mass loss rates predicted by the scaling law given in Eqn \ref{eqn:1Dscale} are shown by the dashed curves.}
\end{figure}

Calculations of fully 2D rotating stars indicate that the situation may be more complicated, and Equation \ref{eqn:1Dscale} may be used incorrectly in some situations.  We find that for models at constant mass, the overall mass loss rates {\it decrease} as the rotation rate increases, as shown in Figure \ref{time} for 20 \msun\ models.  As the rotation rate increases, the effective temperature and luminosity also decrease, as shown in Table \ref{models}, which leads to lower mass loss rates in all cases.  The decrease is about 0.06 dex for the K89 rates, and about 0.15 dex for the V01 rates.  At 30 and 40 \msun, the decrease is between 0.03 - 0.05 dex for all three rates, although these models do not extend to as rapid rotation as the 20\msun\ models.   

As discussed above, the scaling law in Eqn \ref{eqn:1Dscale} applies to stars at \emph{constant luminosity and effective temperature}, but is often applied to models with the same mass at different rotation rates even though the rotation changes T$_{eff}$ and L.  The mass loss rates discussed here are held at constant mass, while the effective temperature and luminosity are allowed to vary.  This makes more sense from a stellar modelling stand point as in most evolution codes, models are made at a given mass, and the effective temperature and luminosity are allowed to vary as the rotation rate changes.  For those who wish to implement a mass dependent scaling law, we have found linear fits to our 2D calculations.  For the V01 models, the mass loss rate can be fit linearly in the mass and the square of the ratio of the surface equatorial rotation velocity to the critical velocity (($v/v_{crit}$)$^2$):
\begin{equation}
\label{eqn:vinkfit}
\dot{M}(v) = \dot{M}(v=0) -0.464*(v/v_{crit})^2 +0.007157(M/M_{\odot}).
\end{equation}
The K89 rates are more complicated.  At any given mass, the mass loss rate can be fit linearly as a function of $(v/v_{crit})^2$, but the slope of the line varies with mass.  With only three points, it is difficult to produce a reliable fit, although the result is clearly not linear.  We present here fits at each mass individually:
\begin{equation}
\label{eqn:cakfit}
\dot{M}(v) = \dot{M}(v=0) + a(v/v_{crit})^2
\end{equation}
where $a$ = -0.148 at 20 \msun, -0.221 at 30 \msun\ and -0.123 at 40 \msun. 

\subsection{Anisotropic mass loss}

As discussed in Section \ref{massresults}, one of the consequences of rotation is to produce a variation in temperature across the surface of the star, which causes the mass loss rates to vary as well.  Based on the CAK mass loss formalism, the mass loss rate at any given colatitude in a rotating star can be written relative to the polar values, $\dot{M_o}$, $F_o$ and $g_o = GM/R^2$:
\begin{equation}
\label{eqn:2Dscale}
\frac{\dot{M}(\theta)}{\dot{M_o}} = \left[\frac{F(\theta)}{F_o}\right]^{1/\alpha}\left[\frac{g_{eff}(\theta)}{g_o}\right]^{1-1/\alpha}
\end{equation}
where $\alpha$ is taken to be independent of latitude and $g_{eff}$ is the centrifugally reduced surface gravity
\begin{equation}
\label{eqn:2Dscaleb}
g_{eff}(\theta) = \frac{GM}{R^2}(1-\Omega^2sin^2\theta)
\end{equation}
where $\Omega^2=V^2_{rot}R/GM$ \cite{owocki96,owocki98}.  Combining this equation with the \citet{vonz} gravity darkening shows that $\dot{M}(\theta)\propto g_{eff}(\theta)$, and hence the mass loss rate decreases towards the equator \cite{owocki96}.  This is expected to produce bipolar outflows around rotating stars.  Using our 2D stellar models, we can test the efficacy of this scaling law approximation.  

\begin{figure}
\includegraphics[width=\linewidth]{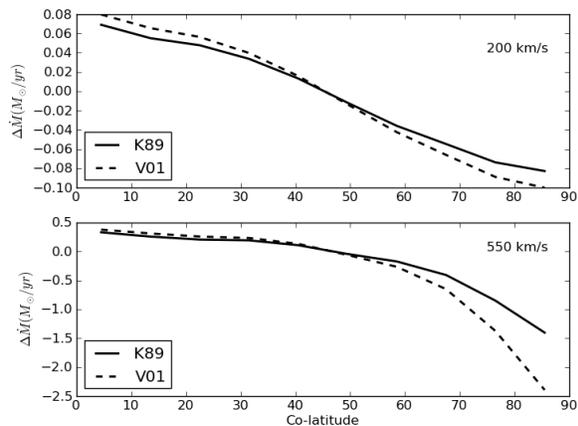}
\caption{\label{fig:20M2D}The relative difference between the 2D calculation and the 1D prediction for mass loss rate from a 20 \msun\ model rotating at 200 (top) and 550 \kms\ (bottom panel)as a function of colatitude for the K89 (solid) and V01 (dashed) mass loss rates.  The 1D prediction is scaled such that the total mass loss is the same.}
\end{figure}

\begin{figure}
\includegraphics[width=\linewidth]{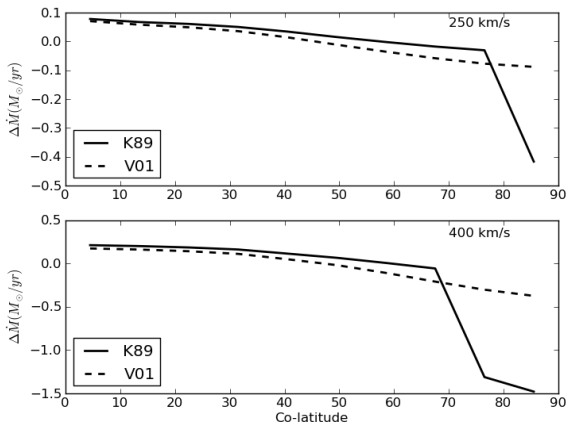}
\caption{\label{fig:40M2D}The relative difference between the 2D calculation and the 1D prediction for mass loss rate from a 40 \msun\ model rotating at 250 (top panel) and 400 \kms\ (bottom panel) as a function of colatitude for the K89 (solid) and V01 (dashed) mass loss rates.  The 1D prediction is scaled such that the total mass loss is the same in both cases.}
\end{figure}

The difference between the 1D predictions and the 2D calculations for mass loss rate across the surface of a rotating star is shown in Figure \ref{fig:20M2D} for a 20 \msun\ model rotating at 200 and 550 \kms and in Figure \ref{fig:40M2D} for a 40 \msun\ model.  We have scaled the mass loss rates predicted by Equation \ref{eqn:2Dscale} so that the total mass loss is the same.  In all cases, the 1D scaling law over predicts the amount of mass loss from the equatorial regions, and under predicts the mass loss from the polar regions.   The differences between the V01 models and the 1D predictions are typically larger than the K89 rates by a factor of 2.  In the most rapidly rotating model (20 \msun, v = 550 \kms), the V01 mass loss rates are 100 \% smaller than the 1D predictions at the equator.  For most models, typical variations for all mass loss rates are between 10-20 \% larger at the pole and smaller at the equator.  

Most of the difference between the 1D and 2D models can be accounted for by differences between the temperatures predicted by von Zeipel's law \cite{vonz} and the temperatures calculated in our models, which are typically in the range of 10 \% \cite{lovekin06}.    The remaining difference arises from the assumptions used in deriving Equation \ref{eqn:2Dscaleb}.  In reality, the force multiplier parameters, $\alpha$, $\delta$ and $k$ are all functions of temperature and luminosity.  In Equation \ref{eqn:2Dscaleb} however, $\alpha$ is taken to be independent of latitude, and only the effective gravity varies across the surface.  In our models, $\alpha$ is nearly independent of temperature, and so we also assume $\alpha$ to be constant.  The second parameter,  $\delta$,  is also approximately constant over the range of temperatures considered here.  However, the third line force parameter, $k$, is strongly dependent on temperature.  This dependence on temperature, which is not accounted for in the 1D scaling relation, will increase the variation between pole and equator in fully 2D mass loss rates.  

\section{Conclusions}
\label{conclusions}

Two different prescriptions for radiatively driven mass loss rates were added to the 2D stellar evolution code {\tt ROTORC} \cite{bob90,bob95}.  We used (1) a modification of the original theoretical description which includes the finite disk approximation by \citet{kud} and (2) mass loss rates based on Monte Carlo simulations by \citet{vink01}.   Either of these mass loss prescriptions can be used to calculate the mass loss using the local properties across the surface of a rotating star, or using the global properties to give a pseudo-1D calculation of the mass loss.  

We considered the total mass lost from rotating stars using our models.  We found that as the rotation rate increases, the mass loss rate decreases for any given mass.  As rotation increases the effective temperature and luminosity of the star decreases, which decreases the mass loss rate.  This effect goes in the opposite direction of the commonly used scaling relation given in Equation \ref{eqn:1Dscale}.  However, this equation is derived, both theoretically and observationally, for stars held at \emph{constant temperature and luminosity}, which is not the case when comparing stars of the same mass and increasing rotation rate.  In Equations \ref{eqn:vinkfit} and \ref{eqn:cakfit}, we provide scaling relations for mass loss rates as a function of rotation rate for stars at constant mass.  

Using our 2D models, we investigated the variation in mass loss rate across the surface of a rotating star.  A direct comparison of our pseudo-1D and 2D calculations show that the total angle integrated amount of mass lost is not significantly affected, but a full 2D calculation dramatically changes the distribution of mass lost.  As our models do not assume von Zeipel's law holds, we can provide an independent verification of the scaling relation based on the CAK theory and von Zeipel's law \citep{vonz} commonly used in 1D models.  We have found that the von Zeipel models predict a smaller range of mass loss rates from the pole to equator than our 2D models, with typical differences of 10-20 \% at the pole and equator.  The difference between the 2D and pseudo-1D model can mostly be explained by differences between our surface temperature structure and that predicted by von Zeipel's law, with the remaining differences a result of allowing the force multiplier parameters to vary with co-latitude in our 2D models.  

The change in distribution of the mass loss also affects the angular momentum loss from these models.  In general, the 2D models predict a lower rate of mass loss at the equator than the pseudo-1D models, for which the mass loss rate is constant across the surface. Although the total rate of angular momentum loss increases as the rotation rate increases, the fraction of the total angular momentum lost is approximately constant for a given mass and mass loss prescription.  However, the scaling changes dramatically with mass and mass loss prescription, so it does not seem that a straightforward scaling relation can be derived.  

As the 2D models presented here lose less angular momentum, they should be expected to spin down more slowly than 1D models.  This could produce one of two effects.  The first possibility is that these models will lose more mass at the pole and remain rapidly rotating through later stages of evolution.  The second possibility is that these models would maintain a rotation rate high enough to reach the critical velocity sooner in the evolution and lose a significant amount of mass (and hence angular momentum) at the equator.  In a future paper, we will investigate how the 2D mass and angular momentum loss affect the evolution of these stars.

\section*{Acknowledgments}
The author would like to thank R. Deupree for helpful discussions and assistance with {\tt ROTORC}, and R. Kudritzki for advice on parameterization of the force multiplier constants.  This work was performed for the U. S. Department of Energy by Los  Alamos National Laboratory under Contract No. DE-AC52-06NA2-5396.

\bsp

\label{lastpage}

\end{document}